\documentclass[%
aip
,10pt,letterpaper,bibnotes,floatfix,raggedbottom,
 amsmath,amssymb,
superscriptaddress
]{revtex4-1}

\usepackage{graphicx}
\usepackage{dcolumn}
\usepackage{bm}

\usepackage[utf8]{inputenc}
\usepackage[T1]{fontenc}
\usepackage{mathptmx}
\usepackage{etoolbox}
\usepackage[dvipsnames]{xcolor}
\usepackage{sidecap}
\usepackage{fancyhdr}

\newcommand*{\citen}[1]{%
  \begingroup
    \romannumeral-`\x 
    \setcitestyle{numbers}%
    \cite{#1}%
  \endgroup   
}

\makeatletter
\def\@email#1#2{%
 \endgroup
 \patchcmd{\titleblock@produce}
  {\frontmatter@RRAPformat}
  {\frontmatter@RRAPformat{\produce@RRAP{*#1\href{mailto:#2}{#2}}}\frontmatter@RRAPformat}
  {}{}
}%
\makeatother
\begin{document}


\pagestyle{fancyplain}
\fancyhf{} 
\renewcommand{\headrulewidth}{0pt}
\fancyhead[R]{\thepage}
\fancyfoot[C]{Distribution A. Approved for public release: distribution unlimited. (AFRL--2024--4036) Date Approved 07--24--2024.}



\title{Extrinsic Dielectric Response due to Domain Wall Motion in Ferroelectric BaTiO$_3$}

\author{Ashok Gurung}
	\email{ashok.gurung@uconn.edu}
	\affiliation{\hbox{Department of Physics, University of Connecticut, Storrs, Connecticut 06269, USA}}
 \author{Mohammad Fatin Ishtiyaq}
	\affiliation{Department of Materials Science \& Engineering, and Institute of Materials Science, University of Connecticut, Storrs, Connecticut 06269, USA}
\author{S.\ Pamir Alpay}
	\affiliation{\hbox{Department of Physics, University of Connecticut, Storrs, Connecticut 06269, USA}}
	\affiliation{Department of Materials Science \& Engineering, and Institute of Materials Science, University of Connecticut, Storrs, Connecticut 06269, USA}
   \author{John Mangeri}
	\affiliation{Center for Atomic-scale Materials Design (CAMD), Department of Physics, Technical University of Denmark, Anker Engelunds Vej 101 2800 Kongens Lyngby, Denmark} 
\author{Serge Nakhmanson}
	\email{serge.nakhmanson@uconn.edu}
	\affiliation{\hbox{Department of Physics, University of Connecticut, Storrs, Connecticut 06269, USA}}
	\affiliation{Department of Materials Science \& Engineering, and Institute of Materials Science, University of Connecticut, Storrs, Connecticut 06269, USA}

\date{\today}

\begin{abstract}
BaTiO$_3$ (BTO) is a prototypical perovskite ferroelectric, whose dielectric permittivity and loss spectra --- 
which are strongly temperature and frequency dependent --- include contributions from inhomogeneous polarization patterns,
with polar domain walls (DWs) being the most common types.
In order to elucidate how DWs influence dielectric response, we utilized a continuum approach based on the Landau-Ginzburg theory
to model field-dependent properties of polydomain tetragonal phase of BTO near room temperature and above.
A system with 180$^\circ$ DWs was evaluated as a case study, with both position-resolved and volume-averaged dielectric susceptibility and loss 
computed at different temperatures for a range of applied field frequencies and amplitudes.
Our results demonstrate that cooperative dipole fluctuations in the vicinity of the DW provide a large (extrinsic) contribution to the system dielectric response relative 
to the intrinsic contribution stemming from within the domain.
Dynamics of the DW profile fluctuations under applied field can be represented by a combination of breathing and sliding vibrational motions,
with each exhibiting distinct dependence on the field frequency and amplitude.
The implications of this behavior are important for understanding and improving control of coupled functional properties in ferroelectric materials, 
including their dielectric, electromechanical, and electrooptical responses. 
Furthermore, this investigation provides a computational benchmark for future studies and can be readily extended to other topological polarization patterns in ferroelectrics.

\end{abstract}

\maketitle

\section{\label{sec:level1} Introduction}

Domain walls (DWs) are ubiquitous in ferroic materials, profoundly influencing their functional behavior \cite{Catalan2012, Schultheiss2023}.
Their topological patterns, mutual orientations, energetics, dynamics and density greatly affect the process of hysteretic switching.
%
%
Furthermore, the same DW characteristics impose an organic sense of length scale on the host system, which in turn gives rise
to a variety of interesting effects that can manifest themselves either with or without an applied field.
Among these effects are large contributions to piezoelectric\cite{BassiriGharb2007, Wada2011, Sluka2012} or piezomagnetic responses\cite{}, 
second-harmonic generation \cite{Bozhevolnyi1998},  anisotropic magnetoresistance\cite{Viret2000, Danneau2002, He2012}, 
unexpected magnetoelectric couplings \cite{He2012, Giraldo2021, Li2023}, enhanced conductivity \cite{Suna2022, Wu2017, Zahn2024}, 
or even onset of superconductivity in adjacent magnetic layers \cite{Yang2004}.
%
%
%

%
In ferroelectrics, DWs form boundary phases between areas with different orientations of spontaneous electric polarization $\mathbf{P}^\mathrm{s}$.
Nanostructured ferroelectrics can exhibit a rich assortment of polarization textures, with rotating components of electric dipoles forming chiral, 
vortex, skyrmionic or other patterns\cite{CherifiHertel2017, Fusil2022, Gregg2012, Das2019}.
The presence of such patterns can underpin novel or unexpected material behavior, e.g., negative capacitance\cite{Zubko2016, Das2021} 
or second-harmonic generation enhancement \cite{Wang2024}.
Strong influence of the system microstructure\cite{Schultheiss2023, Wolk2024}, as well as applied electric, mechanical, thermal and other conditions 
on the shape and evolution of the polarization texture can provide even more levers for controlling and fine-tuning the system properties
and performance.
A characteristic of paramount importance for ferroelectric compounds (as well as most electroactive materials in general)
is the system \emph{small-signal} dielectric susceptibility, i.e., a property that involves fluctuations $\Delta\mathbf{P}$ about the 
equilibrium $\mathbf{P}^\mathrm{s}$ under a weak (subswitching or subcoercive) oscillating electric field $\mathbf{E}$.
Since the early 1960s, researchers have attempted to identify the influence of DWs on the system dielectric response\cite{Fousek1965, Fousek1966, Benguigui1973, Arlt1987, Herbiet1989, BassiriGharb2007, Wang2007, Karthik2012, Xu2014, Fancher2017, Schultheiss2023}.
For weak applied fields, the fluctuations of polarization are small and DWs are assumed to be vibrating. 
Such \emph{extrinsic} contribution to the system small-signal permittivity can usually be measured by applying a bias voltage that causes the underlying domain patterns 
to experience irreversible transformations due to nonlinear switching \cite{Arlt1987, Karthik2012, Xu2014}.
%

%

%
It is expected that this extrinsic contribution associated with the DW presence can comprise a significant portion of the total dielectric response 
signal\cite{Xu2014, Liu2017, Fancher2017, Schultheiss2023}.
Since typical ferroelectric samples host a range of different DW orientations and densities (as well as a variety of different grain and grain-boundary shapes and orientations 
in the case of polycrystalline ceramics\cite{Marincel2015, Schultheiss2023, Wolk2024}), a unified theoretical description of dielectric response in such complicated structures
remains elusive --- especially so for atomistic-scale methodologies that would require large supercells and approximations that describe finite temperature behavior.
Nonetheless, density functional theory and molecular dynamics simulations by Liu and Cohen\cite{Liu2017} showed that polar dipoles at 90$^\circ$ and 180$^\circ$ DWs 
in tetragonal PbTiO$_3$ at room temperature are much more responsive to applied fields, relative to dipoles located deep within the domain.
By parsing out the local contribution due to the DW, these authors demonstrated that its dielectric susceptibility can exceed the intrinsic monodomain one by 2--6 times.
This result agrees well with the earlier theoretical models proposed by Fousek\cite{Fousek1965} and Arlt \emph{et al}\cite{Arlt1987} 
based on a quasiparticle approximation of rigid motion of the DW plane.
Recent observations from position-resolved scanning impedance microscopy\cite{Wu2017} suggest that vibrational DW motion in ferroelectric 
hexagonal manganites can be regarded as a \emph{sliding} displacement of a `rigid' DW profile combined with a (lower energy) \emph{breathing} oscillation 
of the profile thickness.
This description of DW dynamics in applied fields is natural, since analogous behavior has long been studied in 
ferromagnets\cite{Slonczewski1984, Stamps1997, Matsushita2014, Mori2014, Metaxas2016}.
However, specific roles such DW sliding and breathing modes play in the extrinsic dielectric response of ferroelectrics are yet to be determined as relatively few 
investigations have been conducted in this area\cite{Jimenez2020, Chen2020, Chen2021}.
%
%
%

%
In this work, we report the results of a comprehensive numerical study of dielectric response of a generic ferroelectric system with DWs utilizing a continuum-scale approach 
based on the Landau-Ginzburg theory.
We have chosen BaTiO$_3$ (BTO) as an example due to an abundance of information on its DW-related properties in bulk, thin film and nanostructural 
forms \cite{tagantsev2010domains,Gregg2012,Ertug2013}, as well as due to an availability of accurate free energy parameterizations for this popular material.
Near room temperature and up to the Curie temperature, $T_\mathrm{C} = 393$ K\cite{Ertug2013}, BTO is tetragonal (space group $P4mm$), permitting six
symmetrically equivalent polar-domain variants with polarization aligned along the $C_4$ axes of the cubic unit cell.
Therefore, in this temperature range, polarization textures including 90$^\circ$ and 180$^\circ$ DWs are most common.
Here, we focused on the 180$^\circ$ DW texture, as in BTO it involves a change in a single component of the $\mathbf{P}^\mathrm{s}$ vector 
between the neighboring domains (i.e., is Ising-like in character\cite{Marton2010}) and it does not require large simulation volumes for conducting calculations.
In the frequency realm, we concentrated on sub-THz bands, which are of interest for microwave electronics applications \cite{Tsurmi2007, Marksz2021}.
At these probing frequencies, (soft) optical phonon resonances are not strongly excited \cite{JonaShiraneBook, Hlinka2008} and, therefore, 
dielectric susceptibility spectra are considered \emph{relaxational} and well represented by underdamped equations of motion\cite{Fousek1965}.
%

%
%
%
For the most interesting case of longitudinal ($\mathbf{E} \, || \, \mathbf{P}$) induced fluctuations, by evaluating \emph{local} or position-dependent dielectric response,
we show that contributions from the DW region to the system dielectric susceptibility are nearly two orders of magnitude larger, compared to the intrinsic contributions from the
`domain proper,' which is in qualitative agreement with previous observations.
We also demonstrate that, as expected, this local enhancement of the dielectric susceptibility is strongly temperature dependent and grows sharply 
as the temperature approaches $T_\mathrm{C}$. 
At room temperature, for characteristic relaxation frequency (i.e., position of the dielectric loss peak) of the monodomain system located at $\sim$500 GHz 
we find the analogous frequency associated with the 180$^\circ$ DW vibration at around 10 GHz.
Furthermore, the DW vibration can be decomposed into a combination of sliding and breathing motions, similar to those observed
in ferromagnets\cite{Slonczewski1984, Stamps1997, Matsushita2014, Mori2014, Metaxas2016}, 
with the two vibrational modes exhibiting distinct dependencies on the external field frequency and amplitude.
From these insights, we can conclude that, in general, extrinsic contributions to the local weak-field dielectric susceptibility from the DWs may deviate appreciably
from those of the `domain proper' (or monodomain bulk), depending sensitively on the DW orientation and type, which in turn stipulate the nature and intensity 
of relaxational processes occurring within the wall under applied field. 
Naturally, in a large sample that may also be polycrystalline, many DWs in multiple different orientations are present, and therefore any single DW contribution to the dielectric response 
is likely to be washed out.
Still, the simple problem considered here that involves a response from a single DW can serve as a useful \emph{benchmark} for evaluating such extrinsic contributions 
to the total dielectric susceptibility during subcoercive electric loading.
The presented modeling approach can be generalized to other ferroic materials (based on availability of appropriate Landau-type free-energy parameterizations\cite{Mangeri2023}), 
other types and orientations of DWs (e.g., for orthorhombic and rhombohedral BTO phases at lower temperatures\cite{Marton2010}), or for studies of nanostructured materials 
displaying more interesting topological polarization ordering patterns \cite{Gregg2012, Karpov2017, CherifiHertel2017, Das2019}.
%
%

%
\section{Methods}\label{sec:methods}
\subsection{Governing equations}\label{subsec:phenom_time}
In the thermodynamic limit, the total free energy $F$ of a ferroelectric system with a volume $\Omega$ can be represented by a linear combination of energy terms 
corresponding to different physical processes and interactions:
\begin{equation}\label{free_energy}
F(T) = \! \int\limits_{\Omega} \! \left[f_\mathrm{bulk}(T)+f_\mathrm{wall}+f_\mathrm{elastic}+f_\mathrm{strictive}\right] d^3\mathbf{r}.
\end{equation}
Here $f_\mathrm{bulk}(T)$ is the temperature-dependent bulk Landau-type expansion of the ferroelectric ordering energy density at temperature $T$, 
$f_\mathrm{wall}$ is the energy density arising from local gradients of the polarization field $\mathbf{P}$ distribution near DWs, 
$f_\mathrm{elastic}$ is the linear elastic energy density, 
and $f_\mathrm{strictive}$ is accounting for the electrostrictive coupling between $\mathbf{P}$ and 
total strain field tensor ${\bm \varepsilon} = \frac{1}{2}\left({\bm \nabla}\cdot\mathbf{u}+({\bm \nabla}\cdot\mathbf{u})^t\right)$, 
with $\mathbf{u}$ being the elastic displacement field.
The specific form and numerical parameterization of the total-free energy expression used in this work is obtained from Ref.~[\citen{Hlinka2006}].
%
%

%
Time evolution of the polarization field $\mathbf{P}(\mathbf{r},t)$ under applied electric field $\mathbf{E}$ was
obtained as a solution of the time-dependent Landau-Ginzburg-Devonshire (TDLGD) equation:
\begin{equation}\label{eq:TDLGD}
- \Gamma \frac{\partial \mathbf{P}}{\partial t} =  \frac{\delta F(T)}{\delta \mathbf{P}} - \mathbf{E},
\end{equation}
where $\Gamma$ is a time-scale coefficient associated with dynamics of $\mathbf{P}$ on the energy surface and the subsequent damping force corresponding to the 
interaction with defects or the finite temperature anharmonic phonon continuum (see Sec.~\ref{subsec:ac} for further discussion of this coefficient).
When $\mathbf{E} \equiv - \nabla \Phi$ is zero, Eq.~(\ref{eq:TDLGD}) is diffusive in nature, with solutions for $\mathbf{P}$ produced by the process of a gradient descent on 
the system energy surface towards a local or global minimum.
However, a nonzero $\mathbf{E}$ provides a driving force that can oppose the action of stochastic forces associated with the variational derivative of $F(T)$,
which can result in underdamped oscillatory behavior of $\mathbf{P}$ for certain amplitudes of (oscillating) $\mathbf{E}$ and choices of $\Gamma$.
For every time step in the evolution of Eq.~(\ref{eq:TDLGD}), the mechanical equilibrium equation ${\bm \nabla} : {\bm \sigma} = 0$ was also solved, 
where ${\bm \sigma} = \mathbf{C}:{\bm \varepsilon}$ is the total elastic stress field tensor and $\mathbf{C}$ is the fourth-rank elastic stiffness tensor.
In addition, we assumed that Eq.~(\ref{eq:TDLGD}) is satisfied electrostatically at every time step by solving the associated Poisson equation $\epsilon \nabla^2 \Phi = - \nabla \cdot \mathbf{P}$.
The isotropic quantity $\epsilon \simeq 10 \, \epsilon_0$ represents a background dielectric constant due to core-electrons, which is a typical approximation employed
in such continuum-scale simulations \cite{Hlinka2006}.
%

%
%
%
%

%
\subsection{Stationary 180$^\circ$ DWs at zero applied field}\label{subsec:DWs}
\begin{figure}[h]
\centering
\includegraphics[height=12cm,width=\textwidth,keepaspectratio]
{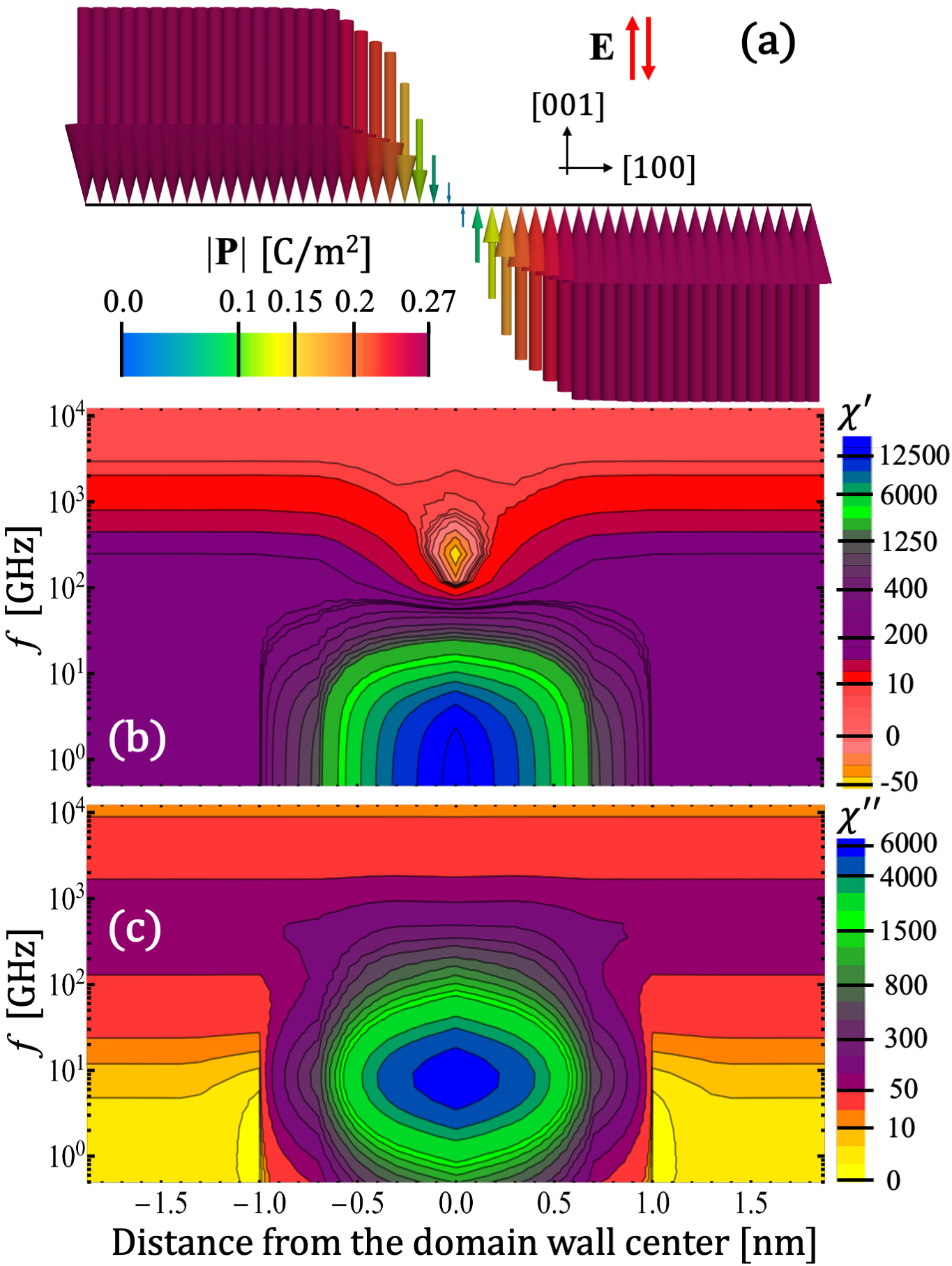}
\caption{
(a) An equilibrium 180$^\circ$ DW profile at RT in the absence of external applied field $\mathbf{E}$.
The \emph{longitudinal} direction of the field application during the dielectric response evaluation is also shown in the upper right corner of the panel.
Contour plots of position-dependent longitudinal ($\mathbf{E}\, || \, P_z \hat{\mathbf{z}}$) dielectric susceptibility at RT as a function 
of the applied field frequency $f$: (b) real part $\chi^\prime (x,f)$ and (c) imaginary part $\chi^{\prime\prime} (x,f)$.
The $x$ axis is common for all the three panels, with its zero mark coincident with the location of the DW profile center, $Pz(x\equiv0) = 0$, in the absence of the applied field.
Similar plots can be obtained for other temperatures.
}
\label{Figure1}
\end{figure}

For the geometrical model of the system volume $\Omega$,
we utilized a quasi-1D rectilinear shape with real-space dimensions of 40 $\times$ 1 $\times$ 1 nm.
The  [100], [010], [001] crystallographic directions of the perovskite BTO unit cell were oriented along the `global' $x,y,z$ cartesian axes of the computational volume.
The governing equations discussed in Sec.~\ref{subsec:phenom_time} were discretized within $\Omega$ utilizing a finite-element method (FEM) approach with hexahedral elements.
Specifically, the \textsc{Ferret} package \cite{Mangeri2017, FerretLink}, based on the open-source Multiphysics Object-Oriented Simulation Environment (MOOSE) framework \cite{permann2020moose}, was used to conduct all the FEM-based simulations described below.
3D periodic boundary conditions were applied to the system field variables $\mathbf{P}(\mathbf{r},t)$ and $\mathbf{u}(\mathbf{r},t)$, leveraging the global strain approach
for the latter\cite{Biswas2020}.
Sine profile initial conditions were chosen for $\mathbf{P}(\mathbf{r},t=0)$, leading to the formation of two $180^\circ$ DWs oriented along the long side of $\Omega$ 
upon energy minimization of Eq.~(\ref{eq:TDLGD}) in zero applied field $\mathbf{E}$.
The 40 nm length of the computational volume was found to be sufficient to treat the neighboring DWs as mostly non-interacting entities.
The system was considered converged when the relative change of $F$ between consecutive time steps fell below $10^{-8}$.
Multiple relaxations of the system with 180$^\circ$ DWs for the zero applied field were conducted for different temperatures, which ranged from room temperature (RT) 
to $\sim$380~K, producing DW profiles with different thickness. 
We found that for elevated temperatures of above $\sim$320~K, discretization meshes with a step of 0.5 nm were optimal for achieving fast convergence,
while for temperatures in the vicinity of RT, due to the DW profiles being more narrow, meshes with a step of 0.333 nm were preferred.
A representative converged ground state of $\mathbf{P}(\mathbf{r})$ at RT is displayed in Fig.~\ref{Figure1}(a), showing one of the two $x$-oriented DWs.
The color map represents the magnitude of the polar vector $|\mathbf{P}|$, which is equivalent to $|P_z|$ for the 180$^\circ$ DW with the chosen orientation.
The DW profile is Ising-like in character, i.e., the up ($+\hat{\mathbf{z}}$) and down ($-\hat{\mathbf{z}}$) oriented polar domains are separated by a region where $|\mathbf{P}|$ tends to zero.
%
%
%

%

\subsection{Evaluation of the dielectric response} \label{subsec:ac}
Under an applied electric field signal of the form of a harmonic wave with angular frequency $\omega$, $\mathbf{E} = \mathbf{E}^{0} e^{i \omega t}$, 
we consider fluctuations of the induced polarization $\Delta \mathbf{P} \equiv \mathbf{P} - \mathbf{P}^\mathrm{s}$.
In case of a weak, subcoercive applied field $|\mathbf{E}^{0}| \ll |\mathbf{E}_\mathrm{C}|$, we can assume that 
$|\Delta \mathbf{P}| \ll |\mathbf{P}^\mathrm{s}|$ and that the shape of the polarization fluctuation signal is also harmonic, but includes a phase delay
$\theta$ with that of $\mathbf{E}$, i.e., $\Delta \mathbf{P} = \Delta \mathbf{P}^{0} e^{i\left(\omega t + \theta\right)}$.
The linear dielectric susceptibility tensor can then be decomposed into real and imaginary components, respectively, as
\begin{equation}\label{eq:real_im_Diel}
\chi^{\prime}_{jk} = \frac{\Delta P_j^{0}}{\epsilon_0 E_k^0}\cos(\theta) \quad \mathrm{and} \quad
\chi^{\prime\prime}_{jk} = \frac{\Delta P_j^{0}}{\epsilon_0 E_k^0}\sin(\theta),
\end{equation}
where $\epsilon_0$ is the vacuum permittivity and $j,k$ are Cartesian indices.
Therefore, by computing and tracking the shape of the $\Delta \mathbf{P}(t)$ signal for a sufficient number of cycles, 
Eqs.~(\ref{eq:real_im_Diel}) can be fit as a function of the applied field frequency $f = \omega /2\pi$, providing information about both 
real and imaginary parts of the system dielectric response. 
The developed methodology presented here and in our previous investigation\cite{BSTO_gurung} allows us to evaluate the response of 
the system polarization to applied electric fields on both local and global levels.
In the first case, the polarization fluctuations are computed at specific points in coordinate space, providing the induced polarization 
as a function of position, $\Delta \mathbf{P}(\mathbf{r})$, which in turn yields position-dependent or \emph{local} estimates for $\chi^\prime_{jk}(f,\mathbf{r})$
and $\chi^{\prime\prime}_{jk}(f,\mathbf{r})$.
In the second case, local $\Delta \mathbf{P}(\mathbf{r})$ fluctuations can be averaged over $\Omega$, 
which includes both domain and DW regions, thus producing volume-averaged or \emph{global} estimates for $\chi^\prime_{jk}(f)$
and $\chi^{\prime\prime}_{jk}(f)$ that lose their dependence on $\mathbf{r}$.
For mapping out the $\chi^\prime$ and $\chi^{\prime\prime}$ frequency dependencies, 
we employed a nonlinear least-squares curve-fitting approach based on the trust-region reflective method, 
which is a robust algorithm for fitting large sparse problems with bounds\cite{trustRegionPython,trustRegion1}.
In this and previous investigation\cite{BSTO_gurung}, we found that the value of the (scalar) time relaxation parameter $\Gamma$ merely influences the time scale of the relaxation, 
while leaving all other features of the dispersion curves (thermal, spatial, or otherwise) intact.
Here, we chose $\Gamma = 3.33\times 10^{-4}$ $\mathrm{kg}\, \mathrm{m}^3 \,\mathrm{s}^{-1} \,\mathrm{C}^{-2}$, 
so that the natural relaxation frequency of the homogeneous monodomain system, 
or equivalently the point when $\chi^\prime (f) = \chi^{\prime\prime} (f)$, is located approximately at 500 GHz. 
%
%
%
%
%
%
%
%
This is a reasonable approximation considering dielectric response measurements provide a similar characteristic relaxation frequency in cubic BTO 
of around 300 GHz\cite{Ponomareva2008}, whereas below $T_\mathrm{C}$ it can be as large as 800 GHz due to mixed order-disorder / displacive nature 
of the phase transition\cite{Hlinka2008}.
For the applied field directions, we probed all three symmetrically inequivalent possibilities, i.e., a longitudinal case of $\mathbf{E}\, || \, [001] \, || \, P_z \hat{\mathbf{z}}$,
as well as both transverse cases $\mathbf{E}\, || \, [100] \, \perp \, P_z \hat{\mathbf{z}}$ and $\mathbf{E}\, || \, [010] \, \perp \, P_z \hat{\mathbf{z}}$.
We found that both transverse modes do not produce DW vibration excitations due to the Ising-like nature of the $z$-axis oriented wall profile. 
%
Therefore, in what follows we present the results associated only with the longitudinal mode DW excitations, i.e., $\chi^{\prime} = \chi_{zz}^{\prime}$ 
and $\chi^{\prime\prime} = \chi_{zz}^{\prime\prime}$, 
which exhibit the most interesting physical behavior.
%

%
%
%
%
%
%
%
%
%

%
\section{Results}\label{sec:res}
We start the discussion of the produced results with an evaluation of the position-dependent longitudinal dielectric susceptibility of the system at RT as a function 
of the applied field frequency $f$, with its real part $\chi^\prime (x,f)$ and imaginary part $\chi^{\prime\prime} (x,f)$,
presented in panels (b) and (c), respectively, of Fig.~\ref{Figure1}.
In panel (b), we clearly observe a large increase in the real part of the dielectric susceptibility near and at the center of the DW, reaching almost 60 times of 
the low-frequency intrinsic (monodomain) response $\chi^\prime_\mathrm{mono}(f \to 0) \approx 200$.
The same behavior is found in panel (c) the imaginary part of the DW-associated dielectric susceptibility, i.e., the extrinsic dielectric loss peak.
For the intrinsic (monodomain) loss peak positioned around 500 GHz [by our choice of the time relaxation parameter $\Gamma$],
the DW-associated loss peak occurs at around 10 GHz.
Interestingly, at the exact center of the DW, the real part of the dielectric susceptibility becomes slightly negative --- as shown in panel (b) by orange-yellow colors --- 
at around 300 GHz, which may indicate the presence of a retardation effect where $\mathbf{E}$ has already switched direction while $\Delta \mathbf{P}$ has not.
Furthermore, in both cases we find that the extent of the local extrinsic contribution to $\chi^\prime$ and $\chi^{\prime\prime}$ is limited in space
to about $\pm 1$ nm away from the DW center, which is a slightly larger region than that is conventionally defined by the DW thickness 
(that is usually obtained by fitting a hyperbolic tangent function to the stationary DW profile\cite{Marton2010}).
\begin{figure}[htpb!]
\centering
\hspace*{-10pt}\includegraphics[width=0.5\textwidth]{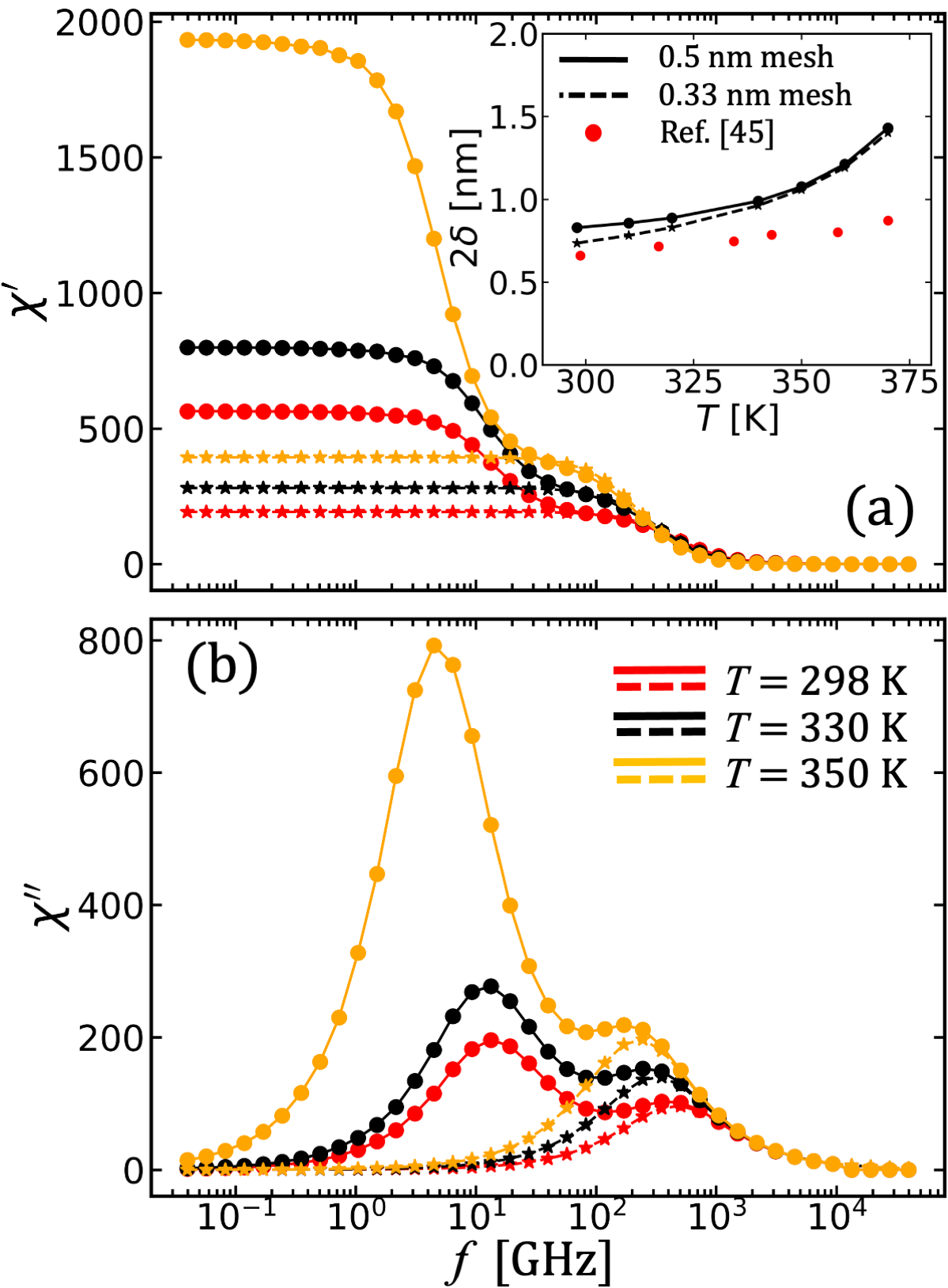}
\caption{
Thermal dependence of the system volume-averaged dielectric susceptibility.
Real (a) and imaginary (b) parts of the dielectric susceptibility [shown in solid lines with circle markers] as a function of frequency
at three different temperatures $T = 298, 330, 350$ K.
The intrinsic contribution is also included in the form of dashed lines with star markers.
The inset depicts the temperature dependence of the computed thickness $2\delta$ of the stationary 180$^\circ$ DW profile. 
Data from Ref.~[\onlinecite{Marton2010}] is provided for comparison.
}
\label{Figure2}
\end{figure}

\begin{figure}[htpb!]
\centering
\hspace{-10pt}\includegraphics[width=0.5\textwidth]{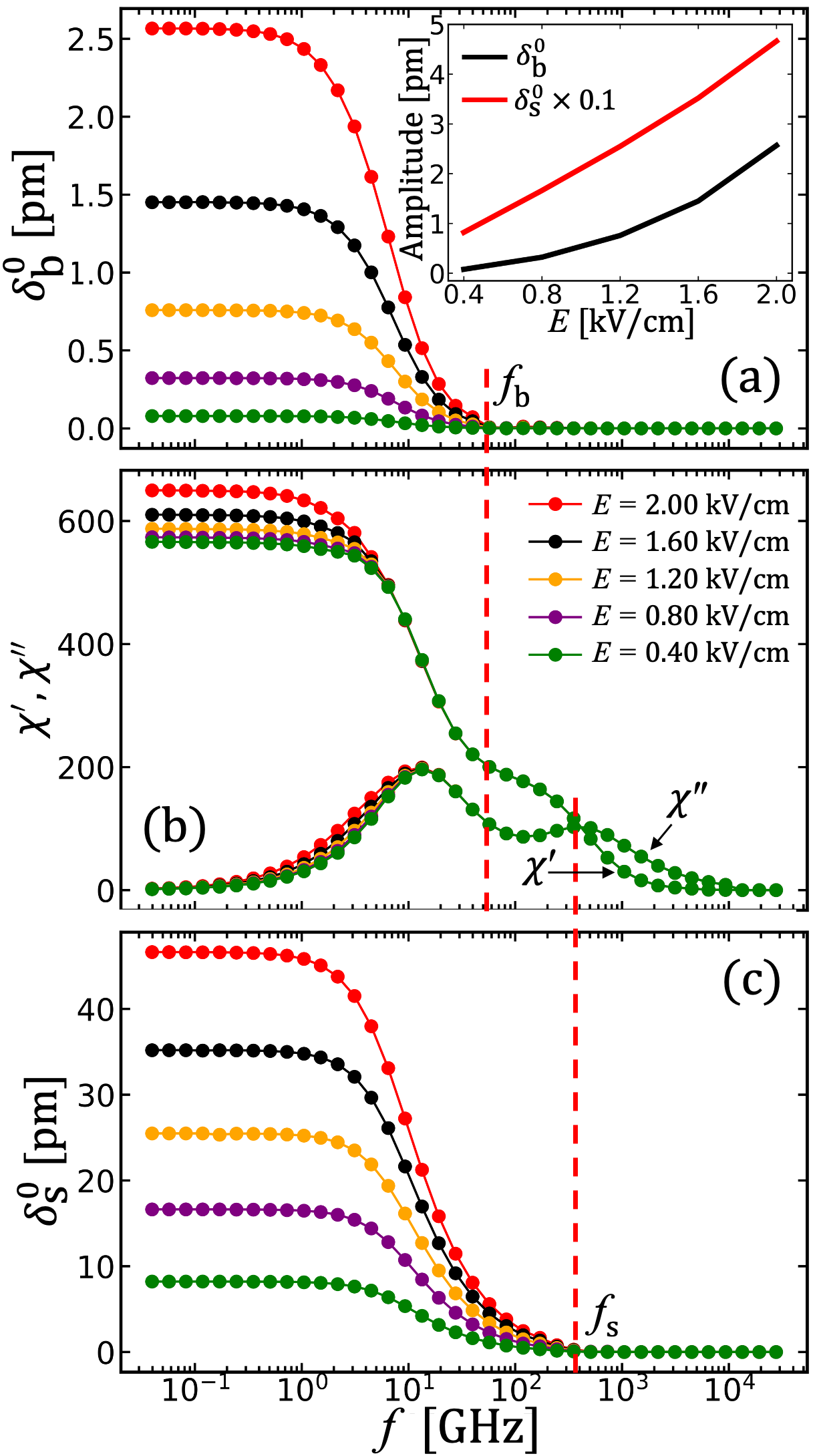}
\caption{
Applied field strength dependence of the system volume-averaged dielectric susceptibility and the DW sliding and breathing vibrational mode amplitudes at RT.
Real [solid lines] and imaginary [dashed lines] parts of the dielectric susceptibility as a function of frequency for a number of different
applied field amplitudes $E \equiv |\mathbf{E}^0|$ are shown in the middle panel (b).
DW breathing mode amplitude $\delta_\mathrm{b}^0$ and sliding mode amplitude $\delta_\mathrm{s}^0$ as a function of frequency for a number of different
applied field amplitudes $E$ are shown in the top panel (a) and bottom panel (c), respectively.
Critical frequencies $f_\mathrm{b}$ and $f_\mathrm{s}$, marking out the complete decay of the amplitude of the corresponding DW vibrational mode,   
are highlighted by dashed red lines.
The inset depicts the dependence of $\delta_\mathrm{b}^0$ and $\delta_\mathrm{s}^0$ on $E$ for $f \to 0$.
}
\label{Figure3}
\end{figure}
Turning to the thermal dependence of the system dielectric response, we present real and imaginary parts of the volume-averaged dielectric susceptibility 
in panels (a) and (b), respectively, of Fig.~\ref{Figure2}.
As the system temperature increases towards $T_\mathrm{C}$, the governing thermodynamic potential $F(T)$ becomes `softer,' 
thus allowing for stronger fluctuations $\Delta \mathbf{P}$ under the applied field. 
Furthermore, as the temperature increases, the DW regions become wider.
In the inset to panel (a), we show the stationary DW thickness $2\delta$ as a function of temperature.
Our calculations for both utilized meshes are in reasonable agreement with results of Ref.~[\onlinecite{Marton2010}] at lower $T$,
but deviate from them towards thicker walls, as the temperature is increased.
E.g., according to our calculations, $\delta( T = 380\,\mathrm{K}) \simeq 2\delta(\mathrm{RT})$.
The increase in DW thickness with growing temperature implies an equivalent expansion in the width of areas associated with the extrinsic dielectric response,
which is reflected in the aggressive enlargement of the respective parts of the volume-averaged dielectric susceptibility curves for both $\chi^\prime$ and $\chi^{\prime\prime}$,
as is shown in Fig.~\ref{Figure2}.
We should mention that, since our simulation volume contains two DWs separated by $20$ nm, these results are presented for a DW volume fraction 
$v_\mathrm{DW} = 2 (2\delta) / 40$.
We find that the volume-averaged $\chi^\prime$ generally follows the rule of mixtures, i.e., it roughly scales as 
$v_\mathrm{DW} \chi^\prime_\mathrm{DW} + (1 - v_\mathrm{DW}) \chi^\prime_\mathrm{mono}$ for any considered $f$, as discussed in Ref.~[\onlinecite{Liu2017}].
Similar aggregate dielectric response curves for subcoercive fields, showing two plateau regions in the $\chi^\prime(f)$ dependence accompanied by two peaks
in the $\chi^{\prime\prime}(f)$ dependence, have been obtained in Ref.~[\onlinecite{Chu2014}] using the phase-field approach 
for the case of generic tetragonal phase 90$^\circ$ DWs. 
The final stage of this study involved a deeper look at the dynamics of DW motion under weak applied electric field,
including an assessment of any dependencies on the field amplitude $|\mathbf{E}^0|$.
Specifically we utilized the following expression to represent the time evolution of the DW profile:
\begin{eqnarray}\label{eq:DW_prof}
P_z(x) = P_z^\mathrm{s} \tanh\left[\delta_\mathrm{b} (x - \delta_\mathrm{s})\right]+\Delta P_z.
\end{eqnarray}
Here, $\delta_\alpha(t), \,\alpha = \mathrm{s,b}$ are the amplitudes of the \emph{sliding} and \emph{breathing} DW vibrational modes, respectively,
described in the same spirit as in the DW dynamics studies of ferromagnetic systems\cite{Slonczewski1984, Stamps1997, Matsushita2014, Mori2014, Metaxas2016}.
For harmonically oscillating $|\Delta \mathbf{P} (t)|$, both amplitudes can also be fitted to harmonic functions:
$\delta_\alpha(t) \equiv \delta_\alpha^0 e^{i (\omega_\alpha t + \theta_\alpha)}$
[note that $\omega_\alpha$, $\theta_\alpha$ can be different from $\omega$, $\theta$].
We illustrate the dependence of the DW breathing and sliding mode amplitudes on frequency in panels (a) and (c) of Fig.~\ref{Figure3}.
This data is obtained for RT, while the applied field strength ranges from 0.4 to 2.0 kV/cm.
We can assign both types of DW motion critical frequencies $f_\alpha$, that are independent of the value of the applied field, beyond which their amplitudes decay to zero.   
In panel (b) of Fig.~\ref{Figure3}, we show how $f_\mathrm{b}$ and $f_\mathrm{s}$ align with the features of the volume-averaged
$\chi^\prime$ and $\chi^{\prime\prime}$ dispersion curves.
Remarkably, each critical frequency corresponds to an inflection point, or a kink, on the $\chi^\prime(f)$ curve, which is a center of a
narrow region where the system `jumps' from one plateau to the next -- revealing exactly how each DW motion contributes to the dielectric
response of the system and where such contribution `shuts off'.
(Obviously, since $f_\mathrm{b} \ll f_\mathrm{s}$, the breathing vibrational mode `shuts off' at much lower frequencies than the sliding mode.) 
Under the growing applied field strength --- up to the limit of 2 kV/cm, above which we no longer observe harmonic $\Delta \mathbf{P}(t)$ oscillations --- 
the amplitudes of sliding and breathing motion increase, but the shape of their dependency on the field magnitude is not the same.
As shown in the inset to Fig.~\ref{Figure3}(a), for the lowest frequencies probed, the breathing mode amplitude has a quadratic dependence on the 
magnitude of the field, whereas the sliding mode dependence is roughly linear. 
We also point out that the values of $\delta_\mathrm{b}^0$ are on the order of a few pm, while those of $\delta_\mathrm{s}^0$ can be tens of pm. 
This result agrees well with empirical estimates dating back to Kittel's Letter on this topic in 1951 \cite{Kittel1951}, where he estimated the 
subswitching sliding mode amplitudes in BTO to be on the order of a few \% of the lattice constant length.
Such a quantitative agreement is usually not expected for Landau-based models but nevertheless we mention it here.
\section{Discussion}
By utilizing underdamped equations of motion for the system polarization field, we have characterized the extrinsic dielectric response of 
180$^\circ$ DWs in BTO in the subcoercive applied field limit.
Our results confirm the current understanding that DW regions provide a large overall contribution to the dielectric susceptibility of the system 
($\approx 60$ times that of the intrinsic response at RT).
The enhancements were localized to the region surrounding the DW that extends slightly beyond the conventional (geometrical) definition of its thickness.
By sampling the thermal dependence of the dielectric response, we revealed that its extrinsic part grows sharply due to the softening of the system 
energy potential as $T \to T_\mathrm{C}$ and the zero-field DW thickness becomes wider.
Finally, we decomposed the DW vibrational motion into sliding and breathing modes, and showed that both of these have unique field frequency and amplitude 
dependencies, thus providing distinct contributions to the response of the system.
%
%
%
Our assumption of underdamped motion of $\mathbf{P}$ implies that it does not exhibit inertial effects where the polarization would 
have dynamics analogous to magnons\cite{Tang2022, Bauer2023}.
Nonetheless, the presented results are mostly valid within the probed frequency interval, while accounting for the inertial effects would constitute a 
next-level approximation to the underlying theory of dielectric response. 
Building on the results of our previous work\cite{BSTO_gurung}, this study is a step forward in understanding the small-signal inhomogeneous dynamics 
of ferroelectric polarization at the continuum level, and it can be readily generalized and extended to a number of other systems and more complex
polar topologies.
For example, in the case of [110]-oriented (or equivalent) 90$^\circ$ DW, its twisting motion (which is not relevant for the longitudinal excitations
evaluated here) should also be considered\cite{Metaxas2016, Juraschek2019}. 
We point out that, upon field cycling, dynamical (twisting) products of $\mathbf{P} \times \partial \mathbf{P} / \partial t$ must be nonzero,
 giving rise to magnetic fields and thus making the DW regions dynamically magnetoelectric\cite{Juraschek2017}.
Corresponding experiments to investigate this effect have already been proposed\cite{Juraschek2019}.

\begin{acknowledgments}
\noindent A.G., S.P.A. and S.N. gratefully acknowledge the Air Force Research Laboratory, Materials and Manufacturing Directorate (AFRL/RXMS) 
for support via Contract No.\ FA8650--21--C5711. 
J.M. thanks Pavel Marton (FZU) for some early stimulating discussions on the topic.
\end{acknowledgments}

\bibliography{ref}

\end{document}